\documentclass[%
aps,
pre,
twocolumn,
reprint,
longbibliography,
preprintnumbers,
amsmath,amssymb,
notitlepage,
floatfix
]{revtex4-1}

\usepackage{siunitx} % units
\usepackage{graphicx}
\usepackage{cancel}
\usepackage{empheq} % boxed eqns, left curly brace eqns
\usepackage{bm} % bold math
\usepackage{mathrsfs}

\usepackage[colorlinks,citecolor=blue]{hyperref}

\newcommand{\rd}{\mathrm{d}}

\DeclareMathOperator{\sgn}{sgn}

\begin{document}

\title{Pressure drop reduction due to coupling between shear-thinning fluid flow and a weakly deformable channel wall: A reciprocal theorem approach}

\author{Shrihari D.\ Pande}
\affiliation{School of Mechanical Engineering, Purdue University, West Lafayette, Indiana 47907, USA}

\author{Ivan C.\ Christov}
\thanks{Corresponding author}
\email{christov@purdue.edu}
\homepage{http://tmnt-lab.org}
\affiliation{School of Mechanical Engineering, Purdue University, West Lafayette, Indiana 47907, USA}

\date{\today}

\begin{abstract}
    We employ the Lorentz reciprocal theorem to derive a closed-form expression for the pressure drop reduction due to the coupling between shear-thinning fluid flow and a weakly deformable channel wall in terms of the shear rate and the viscosity function (and its derivative) of the underlying rigid-channel flow. The methodology is applied in parallel to fluids for which the generalized Newtonian viscosity depends on either the shear rate or the shear stress magnitude. When the viscosity model allows for a closed-form solution for the axial velocity profile in a straight and rigid channel, the pressure drop reduction can be evaluated in closed form, which we demonstrate for the power-law and Ellis viscosity models as featured examples and to enable comparisons to previous works. Importantly, the pressure drop reduction under the Ellis model is valid for both small and large Carreau (or Ellis) numbers, and we show that it reduces to the analytical expression under the power-law model for large Carreau (small Ellis) numbers.
\end{abstract}

\maketitle

% keywords: fluid--structure interaction; shear thinning; reciprocal theorem; lubrication theory; pressure drop

%%%%%%%%%%%%%%%%%%%%%%%%%%%%%%%%%%%%%%%%%%%%%%%%%%%%%%%%%%%%%%%%%%%%%%%%%%%%%%%%%%%%%%%%

\section{Introduction}
Non-Newtonian fluid flows are widely encountered in microfluidic applications, in which complex fluids such as colloidal dispersions and solutions of polymers or nucleic acids flow through long and narrow conduits \cite{Chakra13_v2,Anna13_v2}. For example, it has been of significant interest to employ microfluidic rheometry to characterize the shear-rate-dependent viscosity of these fluids \cite{PM09,GWV16}. However, for typical PDMS-based microfluidic devices \cite{MW02,SW03}, the flow conduits can be compliant, and the channel walls can be deformed by the hydrodynamic pressure of the flow within \cite{GEGJ06,DGNM16,RCDC18}. Understanding how the two-way coupling between the channel's compliance and the fluid's rheology sets the hydrodynamic pressure distribution (and, importantly, the flow rate--pressure drop relation) is key for, e.g., microfluidic design and fabrication \cite{C21}, optimizing fluid-infused laminates for impact mitigation \cite{RHODP24}, actuation of joints in natural and soft-robotic systems \cite{BBG17,GAKZBSS21}, evaluating non-Newtonian lubricants for tribology \cite{AB23,SPCB24}, and Darcy-scale modeling of shear-thinning fluid flows through deformable porous media \cite{RCG23}.
%, to name a few recent applications.

\citet{Y12} was one of the first to consider the flow of a non-Newtonian fluid with shear-dependent viscosity in an axisymmetric compliant tube treated as a Winkler foundation, which presents a two-dimensional (2D) flow problem. He derived a nonlinear ordinary differential equation (ODE) for the tube radius. He examined the ODE's mathematical properties, including the stability of solutions, for both shear-thinning and shear-thickening fluids under a power-law viscosity model. Motivated by microfluidic experiments in three-dimensional (3D) compliant channels, \citet{ADC18} derived the nonlinear ODE for the pressure distribution due to non-Newtonian fluid flow in a compliant, slender, and shallow deformable 3D channel. Like \citet{Y12}, they captured shear-thinning using the power-law viscosity model. Specifically, \citet{ADC18} demonstrated quantitatively that \emph{both} shear-thinning of the fluid and compliance of the channel wall reduce the pressure drop (compared to Newtonian fluid flow in a rigid channel).  \citet{CBCF24} verified these predictions by precision experiments. 

\citet{RAB21} extended the work of \citet{ADC18} by positing that the non-Newtonian fluid obeys a so-called ``simplified'' Phan-Thien--Tanner (sPTT) viscoelastic rheological model (see \cite{DZOP22}, Sec.~III). From the coupled elastohydrodynamic problem, they also obtained a nonlinear ODE for the pressure and solved it numerically to obtain the pressure drop for a fixed flow rate, further quantifying the interplay between compliance and non-Newtonian rheology. An unsteady version of the latter problem was analyzed by \citet{VAN22} to understand how compliance, viscoelasticity, and shear-thinning set the timescale of evolution of peeling fronts. 

However, the sPTT model features both elasticity and shear-thinning and reduces to a Newtonian fluid when either effect is neglected. To disentangle the contributions of elasticity and shear-thinning in compliant channels, a Boger fluid should be considered \cite{J09b}. To this end, \citet{BC22} revisited the problem of a viscoelastic fluid flow in a 3D compliant, slender, and shallow channel. They derived the nonlinear ODE for the pressure distribution under a constant-viscosity viscoelastic Oldroyd-B rheological model. Specifically, they demonstrated a further viscoelastic reduction in the pressure drop (beyond the well-known one due to compliance of the wall) at constant shear viscosity.

Importantly, in a number of previous works (especially on 3D channels), it was not possible to obtain an analytical $q-\Delta p$ relation for arbitrary compliance of the channel wall because of the nonlinear coupling between structural mechanics and fluid mechanics. Nevertheless, in these works, the reduction of the \emph{full} elastohydrodynamic problem to a \emph{single} ODE for the hydrodynamic pressure was a major accomplishment. Of course, perturbative solutions of the ODE could be obtained to address the leading-order effect of flow-induced wall deformation for weakly compliant channels. In the present work, we would like to further focus on this aspect and ask: \emph{What could we learn about non-Newtonian fluid--structure interactions from the textbook solutions for shear-thinning fluid flows in straight, rigid channels?}

To answer the last question, we present an approach for calculating the pressure drop reduction due to coupling between shear-thinning fluid flow and a deformable channel wall via the Lorentz reciprocal theorem \cite{Lorentz_in_book_v2}. Our goal is to show how to implement the reciprocal theorem's application to this problem. Specifically, we consider a 2D channel with a compliant top wall to demonstrate the approach concisely. (The extension to 3D can be accomplished along the lines of Refs.~\cite{BSC22,BC22}.) 

Assuming a \emph{weakly} deformable channel wall and employing domain perturbation, in Sec.~\ref{sec:domain_perturbation} we derive a new statement of the reciprocal theorem for the flow of a generalized Newtonian fluid (with either a shear-rate or shear-stress dependent viscosity function) in a compliant channel. Then, in Sec.~\ref{sec:lubrication}, the reciprocity relation is reduced under the lubrication approximation to yield an \emph{analytical} expression for the pressure drop reduction due to coupling between shear-thinning fluid flow and the weakly deformable wall. To demonstrate the utility of this result, in Sec.~\ref{sec:results}, we consider two prototypical shear-thinning generalized Newtonian viscosity models: power-law and Ellis. Importantly, the calculation under the Ellis model allows us to capture the transition from Newtonian to shear-thinning behavior as a function of the shear rate, unlike previous works that focused on pure shear-thinning in the power-law regime. The analytical results obtained via the reciprocal theorem are shown to agree with the perturbation expansion of the coupled elastohydrodynamic problem. Conclusions and perspectives for future work are summarized in Sec.~\ref{sec:conclusion}.

%%%%%%%%%%%%%%%%%%%%%%%%%%%%%%%%%%%%%%%%%%%%%%%%%%%%%%%%%%%%%%%%%%%%%%%%%%%%%%%%%%%%%%%%

\section{Domain perturbation problem formulation}
\label{sec:domain_perturbation}

\begin{figure}[t]
    \centering
    \includegraphics[width=\columnwidth]{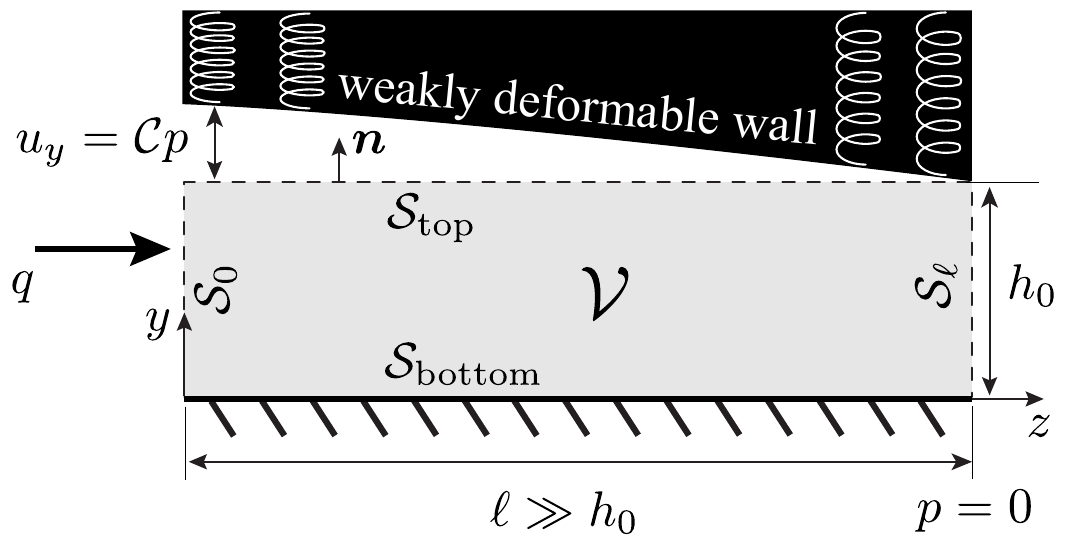}
    \caption{Schematic illustration of the 2D weakly deformable channel geometry. The initially rectangular domain $\mathcal{V}$, 
    of height $h_0$ and axial length $\ell$, over which the reciprocal theorem is applied, is shaded. The surfaces, $\mathcal{S}$, employed in the reciprocal theorem are labeled, along with the outward unit vector $\boldsymbol{n}$. The fluid--solid interface is displaced upwards by $u_y$, which is proportional to the hydrodynamic pressure $p$. }
    \label{fig:channel}
\end{figure}

Consider an initially rectangular fluid domain $\mathcal{V} = \{ (z,y) \,|\, 0\le z\le \ell, \, 0\le y \le h_0 \}$, as shown in Fig.~\ref{fig:channel}. The top wall deforms due to the hydrodynamics forces, and the fluid--solid interface, initially at $y=h_0$, moves to $y=h_0 + u_y$, where $u_y$ is its vertical displacement. Let $\beta = \mathrm{O}(u_y)/h_0$ be the \emph{compliance number}, which quantifies the ``strength'' of fluid--structure interaction; in other words, $\beta$ controls how deformed the initially rectangular fluid domain has become. To enable the use of the reciprocal theorem in a \emph{weakly} deformable channel ($\beta\ll1$), we proceed with the domain perturbation \cite{VD75,L82} approach introduced by \citet{BSC22} for this problem. 

To this end, we expand the fluid's velocity field $\boldsymbol{v}$, the infinitesimal rate-of-strain tensor $\boldsymbol{E} = (\boldsymbol{\nabla}\boldsymbol{v}+\boldsymbol{\nabla}\boldsymbol{v}^\top)/2$, the hydrodynamic pressure $p$, and the Cauchy stress tensor $\boldsymbol{\sigma}$ into perturbation series in $\beta\ll1$:
\begin{subequations}\label{Expansion}\begin{align}
    \boldsymbol{v} &= \hat{\boldsymbol{v}} + \beta\boldsymbol{v}_{1} + \mathrm{O}(\beta^{2}),\label{Velocity beta expand}\\
    \boldsymbol{E} &= \hat{\boldsymbol{E}} + \beta\boldsymbol{E}_{1} + \mathrm{O}(\beta^{2}).\label{Rate of strain beta expand}\\    
    p &= \hat{p} + \beta p_{1} + \mathrm{O}(\beta^2),
    \label{Pressure beta expand}\\
    \boldsymbol{\sigma} &= \hat{\boldsymbol{\sigma}} + \beta\boldsymbol{\sigma}_1 + \mathrm{O}(\beta^{2}),\label{Stress beta expand}
\end{align}\end{subequations}
Here, the quantities designated by ``hats'' correspond to the problem in a 2D straight and rigid channel, i.e., in a \emph{rectangular domain} (no deformation, $\beta=0$). The quantities subscribed by ``$1$'' are the first-order corrections to the latter due to the deformation of the initially rectangular channel's top wall. The remaining calculations are thus all accurate to $\mathrm{O}(\beta)$.  While performing the perturbation expansion in dimensional variables may be perilous (but not necessarily problematic), it leads to certain pedagogical advantages that will become clear below. Mathematically, this approach is valid because we have essentially constructed an expansion such that, for any field $\boldsymbol{\xi}$, the quantity $\|\boldsymbol{\xi}_1\|/\|\hat{\boldsymbol{\xi}}\|$ is dimensionless and independent of $\beta$.

As is standard, we split the pressure $p$ from the viscous stress $\boldsymbol{\tau}$ and write the Cauchy stress tensor for the fluid as $\boldsymbol{\sigma}  = -p \boldsymbol{I} + \boldsymbol{\tau}$, where $\boldsymbol{I}$ is the identity tensor. Now, shear thinning in the steady flow of a fluid with negligible viscoelasticity can be captured by the \emph{generalized Newtonian fluid} model \cite{BAH87,CR08book}, writing the viscous stress tensor as $\boldsymbol{\tau} = 2\eta \boldsymbol{E}$, where $\eta=\eta(\dot{\gamma})$ or $\eta=\eta(\tau)$ is \emph{not} constant but depends on the magnitude of strain-rate tensor or the magnitude of the shear stress tensor itself, which are defined $\dot{\gamma} = \sqrt{2\boldsymbol{E}\boldsymbol{:}\boldsymbol{E}}$ and $\tau = \sqrt{(\boldsymbol{\tau}\boldsymbol{:}\boldsymbol{\tau})/2}$. (In unidirectional shear flow, $\dot{\gamma}$ and $\tau$ are simply the nontrivial components of $\boldsymbol{E}$ and $\boldsymbol{\tau}$, respectively.) Thus, we must now acknowledge the consequences of the perturbation expansion~\eqref{Expansion} on these quantities.

\subsection{Generalized Newtonian fluid with  \texorpdfstring{$\eta = \eta(\dot{\gamma})$}{eta = eta(dot{gamma})}}

Substituting Eq.~\eqref{Rate of strain beta expand} into the definition of the shear rate and using Taylor series expansions for $\beta\ll1$, we find:
\begin{equation}
    \begin{aligned}
    \dot\gamma &= \sqrt{2\boldsymbol{E}\boldsymbol{:}\boldsymbol{E}} \\
    &= \sqrt{2\hat{\boldsymbol{E}}\boldsymbol{:}\hat{\boldsymbol{E}} + 4\beta\hat{\boldsymbol{E}}\boldsymbol{:}\boldsymbol{E}_1 + \mathrm{O}(\beta^2)} \\
    &= \sqrt{2\hat{\boldsymbol{E}}\boldsymbol{:}\hat{\boldsymbol{E}}} + 2\beta\hat{\boldsymbol{E}}\boldsymbol{:}\boldsymbol{E}_1/\sqrt{2\hat{\boldsymbol{E}}\boldsymbol{:}\hat{\boldsymbol{E}}} + \mathrm{O}(\beta^2) \\
    &= \hat{\dot{\gamma}} + \beta\underbrace{2\hat{\boldsymbol{E}}\boldsymbol{:}\boldsymbol{E}_1/\hat{\dot{\gamma}}}_{\dot{\gamma}_1} + \mathrm{O}(\beta^2),
    \end{aligned}
\end{equation}
whence
\begin{equation}
\begin{aligned}
    \eta(\dot{\gamma}) &= \eta\left(\hat{\dot{\gamma}} + 2\beta\hat{\boldsymbol{E}}\boldsymbol{:}\boldsymbol{E}_1/\hat{\dot{\gamma}} + \mathrm{O}(\beta^2)\right)\\
    &=\hat{\eta} + \beta \underbrace{2\eta'(\hat{\dot{\gamma}}) \hat{\boldsymbol{E}}\boldsymbol{:}\boldsymbol{E}_1/\hat{\dot{\gamma}}}_{\eta_1} \, +\, \mathrm{O}(\beta^2),
\end{aligned}    
\label{eq:eta_expand_dg}
\end{equation}
where $\hat{\eta} = \eta(\hat{\dot{\gamma}})$, $\eta' = \rd \eta/\rd \dot{\gamma}$, and $\eta_1$ is a scalar.

\subsection{Generalized Newtonian fluid with  \texorpdfstring{$\eta = \eta(\tau)$}{eta = eta(tau)}}

Now, we start from a perturbation expansion of the shear stress magnitude in terms of the compliance number:
\begin{equation}
    \tau = \hat{\tau} + \beta \tau_1 + \mathrm{O}(\beta^2).
\end{equation}
Then,
\begin{equation}
\begin{aligned}
    \eta(\tau) &= \eta\left( \hat{\tau} + \beta \tau_1 + \mathrm{O}(\beta^2) \right)\\
    &=\hat{\eta} + \beta \underbrace{\tau_1 \eta'(\hat{\tau}) }_{\eta_1} \, +\, \mathrm{O}(\beta^2),
\end{aligned}
\label{eq:eta_expand_dt}
\end{equation}
where $\hat{\eta} = \eta(\hat{\tau})$, $\eta' = \rd \eta/\rd \tau$, and $\eta_1$ is another scalar.

Next, we seek to eliminate $\tau_1$ from the definition of $\eta_1$ arising from Eq.~\eqref{eq:eta_expand_dt}. To this end, we first compute $\tau$ by forming $\boldsymbol{\tau}\boldsymbol{:}\boldsymbol{\tau}$ explicitly from the constitutive relation, $\boldsymbol{\tau} = 2 \eta(\tau) \boldsymbol{E}$. Since $\eta > 0$, it follows that $\tau = \eta(\tau)\dot{\gamma}$. Next, performing the domain-perturbation expansion, we obtain 
\begin{equation}
\begin{aligned}
    \tau = \eta(\tau)\dot{\gamma} &= \eta\left(\hat{\tau} + \beta \tau_1 +   \mathrm{O}(\beta^2) \right) \left(\hat{\dot{\gamma}}_1 + \beta\dot{\gamma}_1 + \mathrm{O}(\beta^2) \right) \\
    &= \hat{\eta} \hat{\dot{\gamma}} + \beta \left[\tau_1 \eta'(\hat{\tau})\hat{\dot{\gamma}} + \eta(\hat{\tau}) \dot{\gamma}_1\right] + \mathrm{O}(\beta^2) \\
    &= \hat{\tau} + \beta \tau_1 + \mathrm{O}(\beta^2).
\end{aligned}     
\end{equation}
Matching terms at each order, we find that
$\hat{\tau} = \hat{\eta} \hat{\dot{\gamma}}$ and $\tau_1 = \tau_1 \eta'(\hat{\tau})\hat{\dot{\gamma}} + \hat{\eta} \dot{\gamma}_1$ or $\tau_1 = \hat{\eta} \dot{\gamma}_1/[1 - \eta'(\hat{\tau})\hat{\dot{\gamma}}]$, which yields
\begin{equation}
    \eta_1 = \frac{\eta'(\hat{\tau}) \hat{\eta} \dot{\gamma}_1}{1 - \eta'(\hat{\tau})\hat{\dot{\gamma}}}. 
    \label{eq:A_eta_expand_tau}
\end{equation}
Observe that this version of $\eta_1$ depends on $\dot{\gamma}_1$ (and, hence, $\boldsymbol{E}_1$) like the expression for $\eta_1$ arising from Eq.~\eqref{eq:eta_expand_dg}.

\subsection{The reciprocal theorem}

To apply the Lorentz reciprocal theorem \cite{Lorentz_in_book_v2}, we must define the ``hat'' and the ``$1$'' (``unknown'') problems' governing equations \cite{HB83,MS19}. At low Reynolds number (negligible flow inertia), the steady flow is governed by the equations of conservation of mass and linear momentum, $\boldsymbol{\nabla} \boldsymbol{\cdot} {\boldsymbol{v}} = 0$ and $\boldsymbol{\nabla}\boldsymbol{\cdot}{\boldsymbol{\sigma}}=\boldsymbol{0}$, respectively, into which the perturbation expansion \eqref{Expansion} is substituted. We refer the reader to, e.g., Ref.~\cite{BSC22} for the details of this straightforward step.

The hat problem at $\mathrm{O}(1)$ corresponds to steady shear-thinning fluid flow in a rigid rectangular channel, and it is
\begin{subequations}\begin{equation}
    \boldsymbol{\nabla} \boldsymbol{\cdot} \hat{\boldsymbol{v}} = 0,\qquad \boldsymbol{\nabla}\boldsymbol{\cdot}\hat{\boldsymbol{\sigma}}=\boldsymbol{0},
    \label{eq:COM_COLM_hat}
\end{equation}
where
\begin{equation}
    \hat{\boldsymbol{\sigma}}=-\hat{p}\boldsymbol{I} + 2\hat{\eta} \hat{{\boldsymbol{E}}} .
    \label{eq:sigmaf_hat}
\end{equation}\end{subequations}

The ``$1$'' problem governs the correction to the hat problem due to fluid--structure interaction between the shear-thinning flow and compliant top wall of the channel, and it is
\begin{subequations}\begin{equation}
    \boldsymbol{\nabla} \boldsymbol{\cdot} \boldsymbol{v}_1=0,\qquad \boldsymbol{\nabla}\boldsymbol{\cdot}\boldsymbol{\sigma}_1=\boldsymbol{0},
    \label{eq:COM_COLM_1}
\end{equation}
where
\begin{equation}
    \boldsymbol{\sigma}_1=-{p}_1\boldsymbol{I} + 2\hat{\eta} \boldsymbol{E}_1 + 2 \eta_1 \hat{\boldsymbol{E}},
    \label{eq:sigmaf_1}
\end{equation}\end{subequations}
which is similar to the representations in  Refs.~\cite{L80,L14,DZEP15,E17,BS21,AN23} for the application of the reciprocal theorem to \emph{weakly} non-Newtonian flows, but without such restriction in the present case.

Now, we are in a position to derive the Lorentz reciprocity relations. To this end, as is standard and shown in several works (e.g., \cite{DS00,MS19,BS21,BSC22}), we multiply the momentum equations from \eqref{eq:COM_COLM_hat} and \eqref{eq:COM_COLM_1} by $\boldsymbol{v}_1$ and $\hat{\boldsymbol{v}}$, respectively and simplify using vector calculus identities and the forms of the stress tensors in Eqs.~\eqref{eq:sigmaf_hat} and \eqref{eq:sigmaf_1}, to obtain:
\begin{subequations}\begin{align}
    \boldsymbol{\nabla}\boldsymbol{\cdot}(\hat{\boldsymbol{\sigma}}\boldsymbol{\cdot}\boldsymbol{v}_{1})-\hat{\boldsymbol{\sigma}}\boldsymbol{:}\boldsymbol{E}_{1} &= 0,\\
    \boldsymbol{\nabla}\boldsymbol{\cdot}(\boldsymbol{\sigma}_1\boldsymbol{\cdot}\hat{\boldsymbol{v}})-\boldsymbol{\sigma}_1\boldsymbol{:}\hat{\boldsymbol{E}} &= 0.    
\end{align}\end{subequations}
Subtracting the first from the second of the latter two equations: 
\begin{equation}
    \boldsymbol{\nabla}\boldsymbol{\cdot} \big(\boldsymbol{\sigma}_1\boldsymbol{\cdot}\hat{\boldsymbol{v}} - \hat{\boldsymbol{\sigma}}\boldsymbol{\cdot}\boldsymbol{v}_{1}\big) = 2\eta_1\hat{\boldsymbol{E}}\boldsymbol{:}{\hat{\boldsymbol{E}}} .
\end{equation}

Finally, applying the divergence theorem over the closed volume $\mathcal{V}$ of the initial, \emph{rectangular} channel (with enclosing surfaces $\mathcal{S}_0$, $\mathcal{S}_\ell$, $\mathcal{S}_\text{top}$, and $\mathcal{S}_\text{bottom}$, with respective outwards unit normals $\boldsymbol{n}$, see Fig.~\ref{fig:channel}), the reciprocal theorem for our problem takes the form:
\begin{multline}
    \underbrace{\iint_{\mathcal{S}_{0}} \big(\boldsymbol{n}\boldsymbol{\cdot}\boldsymbol{\sigma}_1\boldsymbol{\cdot}\hat{\boldsymbol{v}}  -\boldsymbol{n}\boldsymbol{\cdot}\hat{\boldsymbol{\sigma}}\boldsymbol{\cdot}\boldsymbol{v}_{1} \big) \,\rd S}_{\text{inlet},~z=0} \\
    +
    \underbrace{\iint_{\mathcal{S}_{\ell}} \big(\boldsymbol{n}\boldsymbol{\cdot}\boldsymbol{\sigma}_1\boldsymbol{\cdot}\hat{\boldsymbol{v}} -\boldsymbol{n}\boldsymbol{\cdot}\hat{\boldsymbol{\sigma}}\boldsymbol{\cdot}\boldsymbol{v}_{1} \big) \,\rd S}_{\text{outlet},~z=\ell}  
    \\
    = \underbrace{\iint_{\mathcal{S}_{\mathrm{top}}}\boldsymbol{n}\boldsymbol{\cdot}\hat{\boldsymbol{\sigma}}\boldsymbol{\cdot}\boldsymbol{v}_{1} \,\rd S}_{\text{deformable wall},~y=h_0}
    \; + \; \underbrace{\iiint_{\mathcal{V}}2\eta_1\hat{\boldsymbol{E}}\boldsymbol{:}\hat{{\boldsymbol{E}}}\,\rd V}_{\text{shear-thinning}},
    \label{eq:RT_0}
\end{multline}
generalizing the results from \cite{BS21,BSC22}. Note that, by no slip, $\hat{\boldsymbol{v}}$ vanishes on $\mathcal{S}_\text{top}$ and $\mathcal{S}_\text{bottom}$, but $\boldsymbol{v}_1$ only vanishes on $\mathcal{S}_\text{bottom}$.
The reciprocal theorem~\eqref{eq:RT_0} holds for both types of viscosity functions considered by using the appropriate definition of $\eta_1$ from either Eq.~\eqref{eq:eta_expand_dg} or Eq.~\eqref{eq:A_eta_expand_tau}.

Now, we wish to apply Eq.~\eqref{eq:RT_0} to long, slender channels with a deformable top wall conveying a shear-thinning fluid without assuming the non-Newtonian contribution (the second term on the right-hand side) is in any way ``small'' or ``weak.''

%%%%%%%%%%%%%%%%%%%%%%%%%%%%%%%%%%%%%%%%%%%%%%%%%%%%%%%%%%%%%%%%%%%%%%%%%%%%%%%%%%%%%%%%

\section{Two-dimensional channel under the lubrication approximation}
\label{sec:lubrication}

We employ the lubrication approximation for a slender flow geometry, namely $\epsilon=h_0/\ell\ll1$ (see, e.g., \cite{L07,S17_LH}). The flow is driven by an imposed flow rate $q$ at the inlet ($z=0$), and a zero-gauge pressure condition, $p=0$, is imposed at the outlet $z=\ell$. The pressure drop, $\Delta p = p(z=0)$ is to be determined. Thus, we make the problem dimensionless for a flow-rate-controlled regime using a velocity scale $v_c=q/(h_0 w)$, where $w$ is the unit width, and a pressure scale $p_c=\eta_0 v_c /(\epsilon h_0)$. Then, let
\begin{subequations}\begin{align}
    Y &= \frac{y}{h_0},\\
    Z &= \frac{z}{\ell},\\
    V_Y(Y,Z) &= \frac{v_y(y,z)}{\epsilon v_c},\\
    V_Z(Y,Z) &= \frac{v_z(y,z)}{v_c},\\
    \dot{\Gamma}(Y,Z) &= \frac{\dot{\gamma}(y,z)}{v_c/h_0},\\
    \mathcal{T}(Y,Z) &= \frac{\tau_{yz}(y,z)}{\eta_0 v_c /h_0},\\
    P(Z) &= \frac{p(z)}{\eta_0 v_c /(\epsilon h_0)},\\
    U(Z) &= \frac{u_y(z)}{u_c},\\
    \mathfrak{H}(\dot{\Gamma}) &= \frac{\eta(\dot{\gamma})}{\eta_0}.
\end{align}\label{eq:nd_vars}\end{subequations}
For two-way-coupled fluid--structure interaction, the typical deformation scale $u_c$ can be expressed as $u_c = \mathcal{C}p_c$, where the compliance (or inverse stiffness) $\mathcal{C}$ has to be determined from the solution of an elasticity problem. Then, considering these dimensionless variables and corresponding scales, we can define the compliance number as $\beta=u_c/h_0=\mathcal{C}p_c/h_{0}\ll1$. 

As reviewed by \citet{C21} and \citet{R24}, various types of 3D elastic responses of the wall can be reduced to a Winkler-foundation-\emph{like} model \cite{DMKBF18} (as conceptualized by the springs in Fig.~\ref{fig:channel}), $u = \mathcal{C} p$, \emph{without assuming} a Winkler foundation from the outset. On the other hand, assuming a 2D configuration from the outset, \citet{SM04} found $\mathcal{C} = b/(2G+\lambda)$, where $b$ is the elastic wall's thickness, $G$ is a shear modulus, and $\lambda$ is Lam\'e's first parameter. \citet{CV20} have pointed out some of the inherent challenges of the 2D assumption, restricting its applicability to sufficiently compressible materials. We will not revisit the solid mechanics problem here but rather build on this extensive literature and its results.

\subsection{Generalized Newtonian fluid with  \texorpdfstring{$\eta = \eta(\dot{\gamma})$}{eta = eta(dot{gamma})}}

Next, we apply the lubrication approximation for a generalized Newtonian fluid, i.e., neglecting terms of $\mathrm{O}(\epsilon)$, to evaluate various terms featured in the reciprocal theorem~\eqref{eq:RT_0}.

The integrals over $\mathcal{S}_0$ and $\mathcal{S}_\ell$ feature products of the form
\begin{equation}
    \left.\boldsymbol{n}\boldsymbol{\cdot}\boldsymbol{\sigma}\boldsymbol{\cdot}{\boldsymbol{v}}\right|_{z=0,\,\ell} \simeq \mp\frac{\eta_0 v_{c}^{2}\ell}{h_{0}^{2}}\left[-P{V}_{Z}\right]_{Z=0,\,1}.
    \label{eq:lubrication_estimates_1}
\end{equation}
Meanwhile, for the integrand of the integral over $\mathcal{S}_\mathrm{top}$, we find
\begin{equation}
    \left.\boldsymbol{n}\boldsymbol{\cdot}\hat{\boldsymbol{\sigma}}\boldsymbol{\cdot}\boldsymbol{v}_{1} \right|_{y=h_0} \simeq \eta_0\frac{v_c^2}{h_0}\left[\mathfrak{H}(\dot{\Gamma})\left(\frac{\partial \hat{V}_Z}{\partial Y}\right)V_{Z,1}\right]_{Y=1}.
    \label{eq:lubrication_estimates_2}
\end{equation}
The shear rate magnitudes are estimated to $\mathrm{O}(\epsilon)$ as
\begin{equation}
    \hat{\dot{\Gamma}} \simeq \frac{\partial \hat{V}_{Z}}{\partial Y},\qquad
    \dot{\Gamma}_1 \simeq \frac{\partial V_{Z,1}}{\partial Y}.
\label{eq:lubrication_estimates_3}
\end{equation}
Similarly, the volume integral's integrand to $\mathrm{O}(\epsilon)$ is
\begin{equation}
    2\eta_1\hat{\boldsymbol{E}}\boldsymbol{:}\hat{{\boldsymbol{E}}} \simeq \eta_0 \frac{v_c^3}{h_0^3} {\frac{h_0}{v_c}}\mathfrak{H}'(\hat{\dot \Gamma})\left(\frac{\partial \hat{V}_Z}{\partial Y}\right)^2\frac{\partial {V}_{Z,1}}{\partial Y}.
    \label{eq:lubrication_estimates_4}
\end{equation}
Note that the factor of $h_0/v_c$ arises from the switch from $\eta'$ to $\mathfrak{H}'$, where the prime still denotes the derivative with respect to the now dimensionless argument.

Employing the estimates, Eqs.~\eqref{eq:lubrication_estimates_1}--\eqref{eq:lubrication_estimates_4}, in the reciprocal theorem \eqref{eq:RT_0} we obtain the dimensionless form of the reciprocal theorem under the lubrication approximation:
\begin{multline}
    \int_{0}^{1}[P_{1}\hat{V}_Z - \hat{P} V_{Z,1}]_{Z=0}\,\rd Y 
    + \int_{0}^{1}[\hat{P} V_{Z,1} - P_1\hat{V}_Z]_{Z=1}\,\rd Y\\
    = \int_0^{1}\left[ \mathfrak{H}(\hat{\dot{\Gamma}})\left(\frac{\partial \hat{V}_Z}{\partial Y}\right)V_{Z,1}\right]_{Y=1}\,\rd Z  \\
    +  \int_0^{1}\int_0^{1} \mathfrak{H}'(\hat{\dot{\Gamma}})\left(\frac{\partial \hat{V}_Z}{\partial Y}\right)^2\frac{\partial {V}_{Z,1}}{\partial Y} \, \rd Y\rd Z .
    \label{eq:RT_deformable_genNewt_dimless}
\end{multline}

Next, we claim that several terms on the right-hand side of Eq.~\eqref{eq:RT_deformable_genNewt_dimless} can be evaluated by manipulating (\emph{but not solving}) the momentum equation under the lubrication approximation \cite{C21}:
\begin{equation}
    \frac{\partial \mathcal{T}}{\partial Y} = \frac{\partial P}{\partial Z}.
    \label{eq:momentum_eq}
\end{equation}
To substantiate our claim, we substitute the perturbation expansion~\eqref{Expansion} into Eq.~\eqref{eq:momentum_eq} to obtain 
\begin{multline}
    \frac{\partial}{\partial Y}\left[\left(\mathfrak{H}(\hat{\dot \Gamma})+\beta\mathfrak{H}'(\hat{\dot \Gamma})\frac{\partial V_{Z,1}}{\partial Y}\right)\left(\frac{\partial \hat{V}_Z}{\partial Y}+\beta \frac{\partial V_{Z,1}}{\partial Y}\right)\right] \\
    =\frac{\partial \hat{P}}{\partial Z}+\beta\frac{\partial P_1}{\partial Z}.
    \label{eq:momentum_eq_expanded}
\end{multline}
Collecting $\mathrm{O}(1)$ terms in Eq.~\eqref{eq:momentum_eq_expanded}, we have
\begin{equation}
    \frac{\partial}{\partial Y}\left(\mathfrak{H}(\hat{\dot \Gamma}) \frac{\partial \hat{V}_{Z}}{\partial Y} \right) = \frac{\partial \hat{P}}{\partial Z}.
    \label{eq:momentum_eq_ord1}
\end{equation}
Integrating Eq.~\eqref{eq:momentum_eq_ord1} from the centerline $Y=1/2$ to $Y=1$ and using the fact that the pressure (under the lubrication approximation) does not vary with $Y$:
\begin{equation}
    \left. \mathfrak{H}(\hat{\dot \Gamma}) \frac{\partial \hat{V}_{Z}}{\partial Y}\right|_{Y=1} = \frac{1}{2}\frac{\partial \hat{P}}{\partial Z},
    \label{eq:hatVZdY_rexpressed}
\end{equation}
where the symmetry condition $(\partial \hat{V}_{Z}/\partial Y) |_{Y=1/2} = 0 $ is applied at the centerline (see, e.g., \cite{ADC18}). Thus, the hat-problem velocity gradient term in the first integral on the right-hand side of Eq.~\eqref{eq:RT_deformable_genNewt_dimless} can be re-expressed in terms of the hat problem's pressure gradient.

Next, collecting $\mathrm{O}(\beta)$ terms in Eq.~\eqref{eq:momentum_eq_expanded}, we have
\begin{equation}
    \frac{\partial}{\partial Y}\left(\mathfrak{H}(\hat{\dot \Gamma}) \frac{\partial V_{Z,1}}{\partial Y} + \mathfrak{H}'(\hat{\dot \Gamma})\frac{\partial V_{Z,1}}{\partial Y} \frac{\partial \hat{V}_Z}{\partial Y}\right)=\frac{\partial P_1}{\partial Z}.
    \label{eq:momentum_eq_ordbeta}
\end{equation}
Integrating Eq.~\eqref{eq:momentum_eq_ordbeta} from the centerline $Y=1/2$ to some arbitrary $Y$ (and, again, using the fact that the pressure does not vary with $Y$):
\begin{equation}
    \left(\mathfrak{H}(\hat{\dot \Gamma}) + \mathfrak{H}'(\hat{\dot \Gamma})\frac{\partial \hat{V}_Z}{\partial Y}\right) \frac{\partial V_{Z,1}}{\partial Y} = (Y-1/2)\frac{\partial P_1}{\partial Z}.
\end{equation}
Thus, the unknown-problem velocity gradient (featured in the second integral on the right-hand side of  Eq.~\eqref{eq:RT_deformable_genNewt_dimless}) can be re-expressed as
\begin{equation}
    \frac{\partial V_{Z,1}}{\partial Y}=(Y-1/2)\frac{\partial P_1}{\partial Z} \left[\frac{1}{\mathfrak{H}(\hat{\dot \Gamma}) + \mathfrak{H}'(\hat{\dot \Gamma})\frac{\partial \hat{V}_Z}{\partial Y}} \right].
    \label{eq:VZ1dY_reexpressed}
\end{equation}
A further simplification arises when the hat problem is for a straight, rigid channel such that $\hat{P}(Z) = \Delta \hat{P}(1-Z)$ and ${\partial \hat{P}}/{\partial Z}=-\Delta \hat{P}$. Substituting the latter, along with Eqs.~\eqref{eq:hatVZdY_rexpressed} and \eqref{eq:VZ1dY_reexpressed} into Eq.~\eqref{eq:RT_deformable_genNewt_dimless}, the reciprocal theorem statement becomes
\begin{multline}
    \int_{0}^{1}[P_{1}\hat{V}_Z - \hat{P} V_{Z,1}]_{Z=0}\,\rd Y + \int_{0}^{1}[\hat{P} V_{Z,1} - P_1\hat{V}_Z]_{Z=1}\,\rd Y\\
    =-\frac{\Delta\hat{P}}{2}\int_0^1  \left. V_{Z,1}\right|_{Y=1}\,\rd Z\\
    + \int_0^1\int_0^{1} \mathfrak{H}'(\hat{\dot{\Gamma}})\left(\frac{\partial \hat{V}_Z}{\partial Y}\right)^2 \frac{\frac{\partial P_1}{\partial Z}(Y-1/2)}{\mathfrak{H}(\hat{\dot \Gamma}) + \mathfrak{H}'(\hat{\dot \Gamma})\frac{\partial \hat{V}_Z}{\partial Y}} \, \rd Y\rd Z.
    \label{eq:RT_deformable_genNewt3}
\end{multline}
Note that, even for a straight, rigid channel, $\hat{V}_Z(Y)$ and $\Delta\hat{P}$ depend on the viscosity model, i.e., on the functional form of $\mathfrak{H}(\dot{\Gamma})$. In general, determining explicit expressions for $\hat{V}_Z(Y)$ and $\Delta\hat{P}$ may be nontrivial \cite{BAH87}.

Finally, $\left. V_{Z,1}\right|_{Y=1}$ is evaluated by domain perturbation \cite{BSC22} as 
\begin{equation}
    \left. V_{Z,1}\right|_{Y=1} = -U(Z) \left. \frac{\partial \hat{V}_Z}{\partial Y} \right|_{Y=1} + \mathrm{O}(\beta).
    \label{eq:Vz1y1_expand}
\end{equation}
Recalling that \emph{both} $\hat{P}$ and $P_1$ are independent of $Y$ and vanish at the outlet ($Z=1$), so that $\Delta \hat{P} = \hat{P}(Z=0)$ and $\Delta P_1 = P_1(Z=0)$, while $\hat{V}_Z$ and $\hat{\dot{\Gamma}}$ are independent of $Z$, and using Eq.~\eqref{eq:Vz1y1_expand}, Eq.~\eqref{eq:RT_deformable_genNewt3} becomes:
\begin{multline}
    \Delta P_{1} \; \overbrace{\int_{0}^{1} \hat{V}_Z|_{Z=0}\,\rd Y}^{\hat{Q}} \,-\, \Delta \hat{P} \overbrace{\int_{0}^{1} V_{Z,1}|_{Z=0}\,\rd Y}^{Q_1}  \\
    = \frac{\Delta\hat{P}}{2} \left. \frac{\partial \hat{V}_Z}{\partial Y}\right|_{Y=1} \int_0^1  U(Z)\,\rd Z\\
    -
    \Delta P_1 \; \int_0^{1} \mathfrak{H}'(\hat{\dot{\Gamma}})\left(\frac{\partial \hat{V}_Z}{\partial Y}\right)^2 \frac{(Y-1/2)}{\mathfrak{H}(\hat{\dot \Gamma}) + \mathfrak{H}'(\hat{\dot \Gamma})\frac{\partial \hat{V}_Z}{\partial Y}} \, \rd Y.
    \label{eq:RT_deformable_genNewt4}
\end{multline}
For a flow-rate-controlled regime, $\hat{Q}=1$ and $Q_1=0$.

As discussed above, many elasticity problems reduce to $U(Z) = P(Z)$ in dimensionless form. Introducing this elastic response model into Eq.~\eqref{eq:RT_deformable_genNewt4}, we obtain
\begin{multline}
    \Delta {P_1} = \frac{\Delta\hat{P}}{2} \left. \frac{\partial \hat{V}_Z}{\partial Y}\right|_{Y=1} \int_0^1 \hat{P}(Z)\,\rd Z \\
    - \Delta P_1 \int_0^{1} \mathfrak{H}'(\hat{\dot{\Gamma}})\left(\frac{\partial \hat{V}_Z}{\partial Y}\right)^2 \frac{(Y-1/2)}{\mathfrak{H}(\hat{\dot{\Gamma}}) + \mathfrak{H}'(\hat{\dot{\Gamma}})\frac{\partial \hat{V}_Z}{\partial Y}} \, \rd Y.
    \label{eq:RT_deformable_genNewt_ND}
\end{multline}
Since the hat problem corresponds to a straight rigid channel, $\int_0^1 \hat{P}(Z) \,\rd Z = \int_0^1 \Delta\hat{P} (1-Z) \,\rd Z = \Delta\hat{P} /2$, which reduces Eq.~\eqref{eq:RT_deformable_genNewt_ND} to
\begin{empheq}[box=\fbox]{align} 
    \Delta {P_1} = \frac{ \frac{1}{4}(\Delta\hat{P})^2 \left. \frac{\partial \hat{V}_Z}{\partial Y}\right|_{Y=1} }{1 + \frac{1}{2} {\displaystyle \int_0^{1}} \mathfrak{H}'(\hat{\dot{\Gamma}}) \hat{\dot{\Gamma}}^2 \frac{2Y-1}{\mathfrak{H}(\hat{\dot{\Gamma}}) + \mathfrak{H}'(\hat{\dot{\Gamma}}) \hat{\dot{\Gamma}}} \, \rd Y}.
    \label{eq:RT_deformable_genNewt_ND1}
\end{empheq}%

Equation~\eqref{eq:RT_deformable_genNewt_ND1} holds for any generalized Newtonian fluid flow through a deformable channel under the lubrication approximation ($\epsilon\ll1$) and for weak wall compliance ($\beta\ll1$). The first restriction will always hold for long and slender channels and pipes, as would be encountered in any fluidic device, while the restriction to weak compliance may be violated for very soft structures or large pressure drops. The two ingredients needed to use Eq.~\eqref{eq:RT_deformable_genNewt_ND1} are 
\begin{enumerate}
    \item the dimensionless generalized viscosity function $\mathfrak{H}(\dot{\Gamma})$ and its derivative with respect to its argument $\mathfrak{H}'(\dot{\Gamma}) \equiv {\rd \mathfrak{H}}/{\rd \dot{\Gamma}}$,
    \item the pressure drop $\Delta \hat{P}$ and the axial velocity profile $\hat{V}_Z(Y)$ in a straight, rigid channel, from which  $\hat{\dot{\Gamma}}(Y) = \partial\hat{V}_Z/\partial Y$ follows.
\end{enumerate}

\subsection{Generalized Newtonian fluid with  \texorpdfstring{$\eta = \eta(\tau)$}{eta = eta(tau)}}

Employing the lubrication estimates in Eq.~\eqref{eq:RT_0}, as in the previous subsection, but now for $\eta=\eta(\hat{\tau})$, the dimensionless form of the reciprocal theorem is
\begin{multline}
    \int_{0}^{1}[P_{1}\hat{V}_Z - \hat{P} V_{Z,1}]_{Z=0}\,\rd Y + \int_{0}^{1}[\hat{P} V_{Z,1} - P_1\hat{V}_Z]_{Z=1}\,\rd Y\\
    =\int_0^1\left[ \mathfrak{H}(\hat{\mathcal{T}})\left(\frac{\partial \hat{V}_Z}{\partial Y}\right)V_{Z,1}\right]_{Y=1}\,\rd Z  \\
    + \int_0^1\int_0^{1}\frac{\mathfrak{H}'(\hat{\mathcal{T}})\mathfrak{H}(\hat{\mathcal{T}})\dot{\Gamma}_1 \hat{\dot{\Gamma}}^2}{1-\mathfrak{H}'(\hat{\mathcal{T}})\hat{\dot{\Gamma}}}  \, \rd Y\rd Z .
    \label{eq:RT_deformable_genNewt_stress}
\end{multline}

The perturbation expansion of the momentum equation~\eqref{eq:momentum_eq} for $\eta=\eta(\hat{\tau})$, at $\mathrm{O}(1)$, yields
\begin{equation}
    \underbrace{ \mathfrak{H}(\hat{\mathcal{T}})\frac{\partial \hat{V}_Z}{\partial Y}}_{\hat{\mathcal{T}}(Y)} = \left(Y-{1}/{2}\right) \frac{\partial \hat{P}}{\partial Z},
    \label{eq:hatVZdY_rexpressed_2}
\end{equation}
similar to Eq.~\eqref{eq:hatVZdY_rexpressed} but written to be valid for any $Y$. At $\mathrm{O}(\beta)$, the momentum equation gives
\begin{equation}
    \frac{\partial \mathcal{T}_1}{\partial Y}=\frac{\partial}{\partial Y}\left(\frac{\mathfrak{H}(\hat{\mathcal{T}}) \dot{\Gamma}_1}{1 - \mathfrak{H}'(\hat{\mathcal{T}})\hat{\dot{\Gamma}}}\right)=\frac{\partial P_1}{\partial Z},
\end{equation}
whence
\begin{equation}
    \dot{\Gamma}_1=\frac{\partial V_{Z,1}}{\partial Y}=(Y-1/2)\frac{\partial P_1}{\partial Z} \left[\frac{1-\mathfrak{H}'(\hat{\mathcal{T}})\hat{\dot{\Gamma}}}{\mathfrak{H}(\hat{\mathcal{T}})}\right],
    \label{eq:VZ1dY_reexpressed_2}
\end{equation}
similar to Eq.~\eqref{eq:VZ1dY_reexpressed}

Finally, using the fact that the hat-problem pressure gradient is constant and substituting Eqs.~\eqref{eq:hatVZdY_rexpressed_2} and \eqref{eq:VZ1dY_reexpressed_2} into Eq.~\eqref{eq:RT_deformable_genNewt_stress}, the reciprocal theorem statement becomes
\begin{empheq}[box=\fbox]{align} 
    \Delta {P_1} = \frac{ \frac{1}{4} (\Delta\hat{P})^2 \left. \frac{\partial \hat{V}_Z}{\partial Y}\right|_{Y=1} }{1 + \frac{1}{2} {\displaystyle\int_0^{1}} \mathfrak{H}'(\hat{\mathcal{T}}) \hat{\dot{\Gamma}}^2 ({2Y-1}) \, \rd Y}.
    \label{eq:RT_deformable_genNewt_ND2}
\end{empheq}
Note that, in addition to the two ingredients needed to use Eq.~\eqref{eq:RT_deformable_genNewt_ND1}, to use Eq.~\eqref{eq:RT_deformable_genNewt_ND2} we also need an expression for $\hat{\mathcal{T}}(Y)$, which is furnished by the right-hand side of Eq.~\eqref{eq:hatVZdY_rexpressed_2}.

%%%%%%%%%%%%%%%%%%%%%%%%%%%%%%%%%%%%%%%%%%%%%%%%%%%%%%%%%%%%%%%%%%%%%%%%%%%%%%%%%%%%%%%%

\section{Illustrated examples and discussion}
\label{sec:results}

Shear thinning is well described by the Carreau viscosity model \cite{Bird76,BAH87}:
\begin{equation}
    \begin{aligned}
    \eta(\dot{\gamma})
    &=\eta_{\infty}+(\eta_{0}-\eta_{\infty})\left(1+|\lambda_r\dot{\gamma}|^{2}\right)^{(n-1)/2}\\
    &=\eta_0\left[\frac{\eta_{\infty}}{\eta_0}+\left(1-\frac{\eta_{\infty}}{\eta_0}\right)\left(1+Cu^2|\dot{\Gamma}|^{2}\right)^{(n-1)/2}\right],
    \end{aligned}
    \label{eq:Carreau}
\end{equation}
having used the shear stress scale from Eq.~\eqref{eq:nd_vars}. 
Here, $\eta_{0}$ and $\eta_{\infty}$ are the zero- and infinite-shear-rate Newtonian plateaus, respectively, $n\in(0,1]$ characterizes shear thinning from $\eta_{0}$ to $\eta_{\infty}$, and $\lambda_r^{-1}$ is the characteristic (``crossover'') shear rate at which shear thinning becomes apparent \cite{DZEP15,CJYMF22}. The viscosity ratio $\eta_\infty/\eta_0$ is small for most shear-thinning fluids and is henceforth neglected ($\eta_\infty/\eta_0\to0$) \cite{CJYMF22,CBCF24}.
%The cases of $n=1$, $\lambda_r=0$, or $\eta_{0}=\eta_{\infty}$ each represent the Newtonian fluid with a constant viscosity $\eta_{0}$.

The Carreau number, $Cu= \lambda_r q/(h_0^2 w)$, represents the ratio of the characteristic shear rate in the channel, due to the imposed flow rate $q$, to the characteristic shear rate $\lambda_r^{-1}$ of the model \cite{SM15,DZEP15,BS21,CJYMF22,CBCF24}. For $Cu\gg1$, it is acceptable to use the power-law viscosity model \cite{Bird76,BAH87}, as demonstrated quantitatively by \citet{CBCF24} for shear-thinning fluid flows in deformable confinements.

\subsection{Power-law viscosity model}
\label{sec:plaw}

The power-law viscosity model, Eq.~\eqref{eq:Carreau} with $\eta_\infty/\eta_0\to0$ and $1\ll Cu^2|\dot{\Gamma}|^2$, is
\begin{equation} 
    \mathfrak{H}(\dot{\Gamma}) = Cu^{n-1} |\dot{\Gamma}|^{n-1},
    \label{eq:plaw}
\end{equation}
whence
\begin{equation}
    \mathfrak{H}'(\dot{\Gamma}) \equiv \frac{\rd \mathfrak{H}}{\rd \dot{\Gamma}} = (n-1)Cu^{n-1} |\dot{\Gamma}|^{n-2} \sgn(\dot{\Gamma}),
\end{equation}
where the ``signum'' function is defined as $\sgn(\xi) = \xi/|\xi|$ for $\xi\ne0$ and $\sgn(0)=0$.
The solution for flow straight, rigid channel, under the viscosity model~\eqref{eq:plaw}, is well known (see, e.g., \cite{CR08book}). Making it dimensionless using the variables from Eq.~\eqref{eq:nd_vars}, we have:
\begin{multline}
    \hat{V}_Z(Y) = \frac{Cu^{1/n-1}}{2^{1+1/n}(1+1/n)}  \left[ 1-|2Y-1|^{1+1/n}\right] 
    (\Delta \hat{P})^{1/n}
    %\left(-\frac{\partial \hat{P}}{\partial Z}\right)^{1/n}
    %\\
    %\times
    %\begin{cases}
    %1-\left(1-2Y\right)^{1+1/n}, &\quad %Y\le 1/2,\\[2mm]
    %1-\left(2Y-1\right)^{1+1/n}, &\quad %Y>1/2,
    %\end{cases}
    \label{eq:rigid_Vz_plaw}
\end{multline}
and
\begin{equation}
    \hat{P}(Z) = \underbrace{Cu^{n-1}2(4+2/n)^{n}}_{\Delta \hat{P}}(1-Z) .
    \label{eq:rigid_DP_plaw}
\end{equation}
So,
\begin{multline}
    \hat{\dot{\Gamma}}(Y) = \frac{\partial \hat{V}_Z}{\partial Y} \\
    = -\frac{Cu^{1/n-1}}{2^{1/n}} |2Y-1|^{1/n} \sgn(2Y-1) (\Delta \hat{P})^{1/n} .
    %\\
    %\times
    %\begin{cases}
    %+\left(1-2Y\right)^{1/n}, &\quad Y\le %1/2,\\[2mm]
    %-\left(2Y-1\right)^{1/n}, &\quad Y>1/2,
    %\end{cases}
    \label{eq:rigid_DG_plaw}
\end{multline}

From Eqs.~\eqref{eq:RT_deformable_genNewt_ND1} and \eqref{eq:rigid_DG_plaw}, we have
\begin{equation}
    \Delta {P_1} =  \frac{-Cu^{1/n-1} \frac{1}{2^{2+1/n}}(\Delta \hat{P})^{1+1/n}}{1 -  Cu^{1/n-1}\frac{(n-1)}{2^{1/n}n(4+2/n)}(\Delta \hat{P})^{1/n} } \Delta\hat{P}.
\end{equation}
Now, substituting $\Delta\hat{P}$ from Eq.~\eqref{eq:rigid_DP_plaw} into the latter, we obtain
\begin{equation}
    \Delta {P_1} = -\frac{1}{2}(2n+1)(\Delta\hat{P})^2.
    \label{eq:RT_DP1_plaw}
\end{equation}

Following \cite{ADC18,CBCF24}, the solution to the coupled elastohydrodynamic problem yields
\begin{equation}
\begin{aligned}
    \Delta P &= \frac{1}{\beta} \left\{ \left[ 1 + \beta (2n+2)\Delta\hat{P} \right]^{1/(2n+2)} - 1 \right\}\\
    &= \Delta\hat{P} \left[ 1 - \tfrac{1}{2}\beta (2n+1) \Delta\hat{P} \right] + \mathrm{O}(\beta^2),
\end{aligned}
\label{eq:DP_plaw_EHD}
\end{equation}
from which we see that $\Delta P_1 = (\Delta P - \Delta \hat{P})/\beta$ is identical to Eq.~\eqref{eq:RT_DP1_plaw} obtained by the reciprocal theorem. 
Observe that the Taylor series in Eq.~\eqref{eq:DP_plaw_EHD} converges only if $|\beta (2n+2) \Delta\hat{P}| < 1$, or $\beta<Cu^{1-n}/[2(4+2/n)^{n}(2n+2)]$. For $n=1$, however, we have $\beta < 1/48 \approx 0.02$, which places a stringent limit on the applicability of the small-$\beta$ (domain perturbation) expansion. For $Cu \gg 1$, the radius of convergence is enlarged, but it is still finite. 

%%%%%%%%%%%%%%%%%%%%%%%%%%%%%%%%%%%%%%%%%%%%%%%%%%%%%%%%%%%%%%%%%%%%%%%%%%%%%%%%%%%%%%%%

\subsection{Ellis viscosity model}
\label{sec:Ellis}

The power-law model~\eqref{eq:plaw} is singular for $Cu\to0$. From a fundamental point of view, it is of interest to capture the $Cu\to0$ transition from power-law shear-thinning behavior back to Newtonian (shear-rate-independent) behavior or vice versa. To this end, we employ the Ellis viscosity model \cite{R60,MB65}. Following \cite{MB65}, the generalized viscosity function is written as
\begin{equation}
    \eta(\tau)=\frac{\eta_0}{1+|\tau/\tau_{1/2}|^{n_e-1}},\qquad \tau_{1/2} > 0.
    \label{eq:Ellis}
\end{equation}
Requiring consistency with the power-law model~\eqref{eq:plaw} derived from the Carreau model~\eqref{eq:Carreau}, we find $n_e = 1/n$ and $\tau_{1/2}=\eta_0\lambda_r^{-1}$ (see also the discussions in Refs.~\cite{C21,CJYMF22}).
Using the shear stress and pressure scales from Eq.~\eqref{eq:nd_vars}, we make Eq.~\eqref{eq:Ellis} dimensionless:
\begin{equation}
    \mathfrak{H}(\mathcal{T}) = \frac{1}{1 + |Cu\mathcal{T}|^{1/n-1}},
\end{equation}
whence
\begin{equation}
    \mathfrak{H}'(\mathcal{T}) \equiv \frac{\rd \mathfrak{H}}{\rd \mathcal{T}} = \frac{-1(1/n-1)|Cu \mathcal{T}|^{1/n-2}Cu\sgn(\mathcal{T})}{(1+|Cu\mathcal{T}|^{1/n-1})^2}.
\end{equation}

Note that, within the context of the Ellis model, it may be more natural to define the Ellis number \cite{PUBP21} $El = \tau_{1/2}h_0/(v_c\eta_0) = Cu^{-1}$. Here, we follow Refs.~\cite{LPDFM22,LMPD22,CJYMF22} in understanding the Ellis model's utility as an approximation of the Carreau model; thus, we prefer to interpret the former's parameters in terms of the latter's and employ the Carreau number $Cu$ in our study.

The axial velocity profile in a straight, rigid channel under the viscosity model~\eqref{eq:Ellis} is also known \cite{S01}. Here, we adopt the solution form of \citet{CLLDF21} (see also \cite{LPDFM22,LMPD22}). Making it dimensionless using the variables from Eq.~\eqref{eq:nd_vars}:
\begin{multline}
    \hat{V}_Z(Y) = \frac{1}{8} \left[1-(2Y-1)^2\right] \Delta\hat{P} \\
    + \frac{Cu^{1/n-1}}{2^{1/n+1}(1+1/n)} 
    \left[1-|2Y-1|^{1/n+1}\right] (\Delta\hat{P})^{1/n}.
    \label{eq:rigid_Vz_Ellis2}
\end{multline}
So,
\begin{multline}
    \hat{\dot{\Gamma}}(Y) = \frac{\partial \hat{V}_Z}{\partial Y} = - \frac{1}{2}  (2Y-1) \Delta\hat{P} \\
    - \frac{Cu^{1/n-1}}{2^{1/n}}  
    |2Y-1|^{1/n} \sgn(2Y-1) (\Delta\hat{P})^{1/n}.
    \label{eq:rigid_dVz_Ellis2}
\end{multline}
Integrating Eq.~\eqref{eq:rigid_Vz_Ellis2} from $Y=0$ to $Y=1$, we obtain the flow rate--pressure drop relation:
\begin{equation}
    \hat{Q} = 1 = \frac{1}{12} \Delta\hat{P} + \frac{Cu^{1/n-1}}{2^{1/n}(4+2/n)} (\Delta\hat{P})^{1/n},
    \label{eq:rigid_DP_Ellis}
\end{equation}
from which $\Delta\hat{P}$ has to be determined as the solution to a nonlinear algebraic equation, in contrast to the explicit result in Eq.~\eqref{eq:rigid_DP_plaw} under the power-law model.

Two limiting cases of Eq.~\eqref{eq:rigid_DP_Ellis} can be easily obtained by neglecting the second term in comparison to the first ($Cu\ll1$, $1/n-1>0$) or the first term in comparison to the second ($Cu\gg1$ ``moderately'' large) on the right-hand side:
\begin{equation}
    \begin{cases}
    \Delta\hat{P} \simeq 12, &\quad Cu\ll1,\\[1mm]
    \displaystyle 
    \Delta \hat{P} \simeq Cu^{n-1}2(4+2/n)^{n}, &\quad Cu\gg1,
    \end{cases}
    \label{eq:rigid_DP_Ellis_limits}
\end{equation}
corresponding to the Newtonian and power-law regime pressure drops.

Substituting the expressions for the velocity profile from Eq.~\eqref{eq:rigid_Vz_Ellis2}, the shear rate from Eq.~\eqref{eq:rigid_dVz_Ellis2}, and the shear stress from Eq.~\eqref{eq:hatVZdY_rexpressed_2}  into the reciprocal theorem result in Eq.~\eqref{eq:RT_deformable_genNewt_ND2}, we obtain the pressure drop reduction:
\begin{equation}
    \Delta {P_1} = -\frac{\frac{1}{4}(\Delta\hat{P})^2\left[\frac{1}{2}\Delta \hat{P} + \frac{1}{2^{1/n}}(\Delta\hat{P})^{1/n}Cu^{1/n-1}\right]}{1+\frac{1-n}{2^{1+1/n}(2n+1)}(\Delta \hat{P})^{1/n} Cu^{1/n-1}} .
    \label{eq:RT_DP1_Ellis}
\end{equation}

Again, two distinguished asymptotic behaviors can be easily obtained from Eq.~\eqref{eq:RT_DP1_Ellis}:
\begin{equation}
    \begin{cases}
    \Delta P_1 \simeq -\frac{1}{8}(\Delta\hat{P})^3 = -216, &\quad Cu\ll1,\\[1mm] 
    \Delta P_1 \simeq -\frac{1}{2}(2n+1)(\Delta\hat{P})^2 \\
    \phantom{\Delta P_1} =
    -Cu^{2n-2} n(4+2/n)^{2n+1}, &\quad Cu\gg1.
    \end{cases}
    \label{eq:RT_DP1_Ellis_limits}
\end{equation}
The first case corresponds to the Newtonian regime, for which we neglect all $Cu$ terms in Eq.~\eqref{eq:RT_DP1_Ellis} for $Cu\ll1$. The second case corresponds to the power-law regime of shear thinning, for which we observe that $(\Delta \hat{P})^{1/n}Cu^{1/n-1}$ in Eq.~\eqref{eq:RT_DP1_Ellis} becomes independent of $Cu$ for $Cu\gg1$ while $\Delta\hat{P} \ll 1$ for $Cu\gg1$, on using Eq.~\eqref{eq:rigid_DP_Ellis_limits}.

\begin{figure}[t]
    \centering
    \includegraphics[width=\columnwidth]{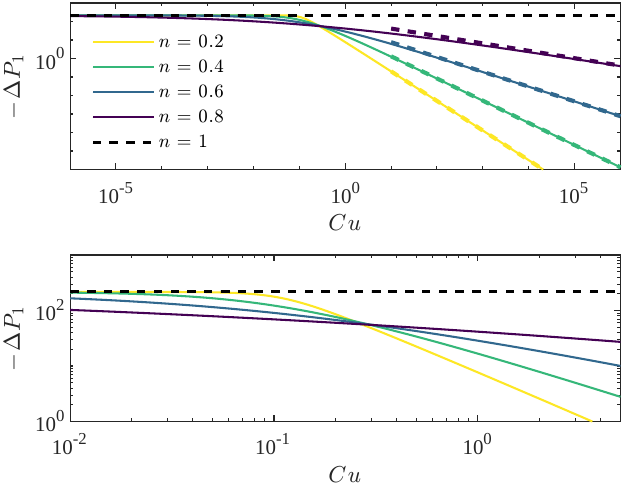}
    \caption{Leading-order correction $\Delta P_1$ to the pressure drop as a function of the Carreau number $Cu$, for different values of the power-law index $n$. The bottom plot is a zoom-in of the top plot to highlight the transition from small to large $Cu$. Solid curves represent the analytical result, Eq.~\eqref{eq:RT_DP1_Ellis}, obtained via domain perturbation and the reciprocal theorem under the Ellis viscosity model, valid for the full range of $Cu$ values shown. Dashed lines represent the analytical result, Eq.~\eqref{eq:RT_DP1_plaw}, obtained via domain perturbation and the reciprocal theorem under the power-law viscosity model, valid for $Cu\gg1$. Black dashed lines represent the corresponding correction to the pressure drop of a Newtonian fluid, see Eq.~\eqref{eq:RT_DP1_Ellis_limits}.}
    \label{fig:dP1_Cu}
\end{figure}

Figure~\ref{fig:dP1_Cu} shows the dependence of $\Delta P_1$ on $Cu$ under the Ellis viscosity model for different values of the power-law index $n$. The pressure drop reduction was calculated by first determining the pressure drop in a rigid channel $\Delta \hat{P}$ from the nonlinear Eq.~\eqref{eq:rigid_DP_Ellis} using  {\sc{Matlab}}'s \texttt{fsolve} and then substituting the numerical value of $\Delta \hat{P}$ into Eq.~\eqref{eq:RT_DP1_Ellis} to obtain the correction due to channel deformation. The figure also shows the solution for $\Delta P_1$ in the power-law regime ($Cu\gg 1$) obtained from Eq.~\eqref{eq:RT_DP1_plaw}. It is evident from the figure that Eq.~\eqref{eq:RT_DP1_Ellis} captures the transition from the Newtonian plateau for $Cu\ll1$ into the power-law regime for $Cu\gg 1$. Importantly, the present approach allows for this smooth transition between Newtonian and shear-thinning behavior in a weakly deformable channel to be described by a closed-form analytical expression, Eq.~\eqref{eq:RT_DP1_plaw}, for which the rigid-channel pressure drop $\Delta \hat{P}$ and the Carreau number $Cu$ are the only inputs. 

To validate our leading-order-in-$\beta$ analytical result, we note that a nonlinear ODE for the pressure under the Ellis model for the coupled elastohydrodynamic problem  (no $\beta\ll1$ assumption) was derived in Ref.~\cite{C21}: 
\begin{multline}
    \left(-\frac{dP}{dZ}\right)^{1/n}(1+\beta P)^{1/n+2}\frac{Cu^{1/n-1}}{2^{1/n+1}(2+1/n)} \\
    -\frac{1}{12}\frac{dP}{dZ}(1+\beta P)^3 = 1.
    \label{eq:Ellis_EHD}
\end{multline}
Equation~\eqref{eq:Ellis_EHD} is an \emph{implicit} differential equation of the form $\mathfrak{F}(Z,P,dP/dZ)=0$, which can be solved numerically using \texttt{ode15i} in {\sc Matlab} \cite{S02_V2}, as also done in Ref.~\cite{RAB21} for the sPTT model and in Refs.~\cite{PUBP21,CJYMF22} for a different problem involving the Ellis model. The latter implements variable-step, variable-order integration based on backward-difference formulas with user-specified tolerances (we set the relative and absolute tolerances to $10^{-12}$ and $10^{-8}$, respectively). Self-consistent initial guesses for the implicit solver were generated from Eq.~\eqref{eq:Ellis_EHD} and the BC $P(1)=0$ using {\sc Matlab}'s \texttt{decic} function. Then, the ODE is ``integrated backward'' towards $Z=0$ (i.e., the implicit ODE is solved under the transformation $Z\mapsto 1-Z$ with ``initial conditions'' at $1-Z = 0$).

In parallel, we can also perform a perturbation expansion of Eq.~\eqref{eq:Ellis_EHD} as $P(Z) = \Delta\hat{P} (1-Z) + \beta P_1(Z) + \mathrm{O}(\beta^2)$, where $\Delta\hat{P}$ satisfies Eq.~\eqref{eq:rigid_DP_Ellis}. Carrying out the expansion to $\mathrm{O}(\beta)$, we find that $P_1(Z)$ satisfies
\begin{multline}
    \left[ \frac{1}{12} + \frac{Cu^{1/n-1}}{2^{1/n+1}(2n+1)} (\Delta\hat{P})^{1/n-1} \right] \frac{dP_1}{dZ} \\
    =  \left[\frac{1}{4} 
    + \frac{Cu^{1/n-1}}{2^{1/n+1}} (\Delta\hat{P})^{1/n-1} \right] (\Delta\hat{P})^2(1-Z),
\end{multline}
which is easily integrated subject to $P_1(1)=0$ to find
\begin{equation}
    \begin{aligned}
        \Delta P_1 &= P_1(0) \\
        &= -\frac{\frac{1}{4}(\Delta\hat{P})^2\left[\frac{1}{2}\Delta\hat{P}+\frac{1}{2^{1/n}} (\Delta\hat{P})^{1/n} Cu^{1/n-1}\right]}{1+\frac{1-n}{2^{1/n+1}(2n+1)} (\Delta\hat{P})^{1/n} Cu^{1/n-1}},
     \end{aligned}
\end{equation}
which is identical to Eq.~\eqref{eq:RT_DP1_Ellis}.

\begin{figure}[h]
    \centering
    \includegraphics[width=\columnwidth]{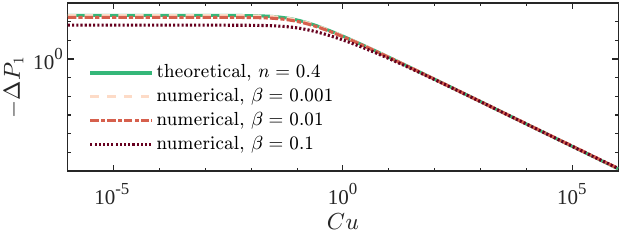}
    \caption{Leading-order correction $\Delta P_1$ to the pressure drop as a function of the Carreau number $Cu$, for different values of the compliance number $\beta$ and a power-law index $n=0.4$. Dashed and dotted curves represent the numerically obtained pressure drop correction from the ODE~\eqref{eq:Ellis_EHD} as $\beta$ is decreased. The solid curve represents the domain-perturbation ($\beta\ll1$) analytical result from Eq.~\eqref{eq:RT_DP1_Ellis}.}
    \label{fig:dP1_beta}
\end{figure}

Figure \ref{fig:dP1_beta} shows the dependence of $\Delta P_1$ on $Cu$ under the Ellis viscosity model for different values of the compliance number $\beta$. The pressure drop correction is obtained by numerically solving the ODE arising from the coupled elastohydrodynamic problem, namely Eq.~\eqref{eq:Ellis_EHD}, for different values of $\beta$. It is evident from the figure that as $\beta\to0$, the numerical solution of Eq.~\eqref{eq:Ellis_EHD} agrees with the leading-order analytical result from Eq.~\eqref{eq:RT_DP1_Ellis}. However, as discussed above, the perturbation series has a small convergence radius. Thus, the value of $\beta$ has to be quite small for the numerical and analytical curves to overlap. Nevertheless, this figure demonstrates the consistency of the domain-perturbation approach for weakly deformable channels and the analytical result for $\Delta P_1$ obtained by the reciprocal theorem in this work. 

%%%%%%%%%%%%%%%%%%%%%%%%%%%%%%%%%%%%%%%%%%%%%%%%%%%%%%%%%%%%%%%%%%%%%%%%%%%%%%%%%%%%%%%%

\section{Conclusion}
\label{sec:conclusion}

In this work, we demonstrated how the Lorentz reciprocal theorem can be used to obtain a closed-form expression for the pressure drop reduction due to shear-thinning fluid flow in a 2D channel with a compliant top wall. The reciprocal-theorem approach offers a different viewpoint on the problem than the domain perturbation approach to the governing differential equations of the elastohydrodynamic problem, solved order-by-order. Specifically, the reciprocal theorem leads us to an expression for the leading-order pressure drop due to compliance in terms of the shear rate and the viscosity function (and its derivative) of the underlying rigid-channel flow. These quantities are of particular interest in non-Newtonian fluid mechanics. On the other hand, the perturbation approach to the governing differential equations requires introducing the viscosity function, $\eta(\dot{\gamma})$ or $\eta(\tau)$, and solving for the flow profile early on in deriving said differential equations, and therefore this approach does not offer a clear view of the role played by the viscosity functions.

The main contribution of our work was to extend the approach of \citet{BSC22} to generalized Newtonian fluids to obtain a reciprocity relation, Eq.~\eqref{eq:RT_0}, for shear-thinning fluid flow in a deformable channel, \emph{without} assuming the non-Newtonian correction is in any way ``small'' or ``weak.'' Employing the lubrication approximation, and manipulations of the momentum equation under the latter, we reduced the reciprocal theorem~\eqref{eq:RT_0} for the cases in which the effective viscosity is a function of the shear rate or the shear stress, obtaining Eqs.~\eqref{eq:RT_deformable_genNewt4} and \eqref{eq:RT_deformable_genNewt_stress}, from which we derived closed-form expression for the (dimensionless) pressure drop reduction $\Delta P_1$ in Eqs.~\eqref{eq:RT_deformable_genNewt_ND1} and \eqref{eq:RT_deformable_genNewt_ND2}, respectively. It should be noted, of course, that though we do not restrict the non-Newtonian contribution to be weak, we do rely on the lubrication approximation to re-express various terms in the reciprocal theorem. In a sense, we trade one difficulty for another but obtain some new insight into the pressure drop reduction. Specifically, $\Delta P_1$ from Eqs.~\eqref{eq:RT_deformable_genNewt_ND1} and \eqref{eq:RT_deformable_genNewt_ND2} can be evaluated even from an easy numerical solution for the flow of a generalized Newtonian fluid \cite{W20} (although, in our featured examples, we used special forms of the viscosity functions that allow for an analytical solution of the axial flow profile).

For the cases of the power-law and Ellis viscosity models, the expressions for $\Delta P_1$ derived from the reciprocal theorem were shown to agree with the perturbation expansions of the respective nonlinear ODEs previously obtained by solving a coupled elastohydrodynamic problem. Importantly, for the Ellis viscosity model, the closed-form expression for $\Delta P_1$ obtained in this work captures the smooth transition between Newtonian behavior and the power-law regime of shear-thinning, as a function of the Carreau number $Cu$, across many orders of magnitude of $Cu$.

In the future, it would be worth considering if a similar analysis could be done for flows in different geometries, such as axisymmetric tubes or 3D channels. Another avenue of future work would be to consider other viscosity models for shear-thinning fluids, for which the solution for the axial velocity profile in a rigid channel is known, such as the special cases of the Carreau model identified by \citet{G20b} or the truncated power-law model of \citet{L15} (see also Ref.~\cite{W20}). Employing these models and their rigid-channel velocity profiles in our main  Eqs.~\eqref{eq:RT_deformable_genNewt_ND1} and \eqref{eq:RT_deformable_genNewt_ND2} could lead to further insight into the coupling between fluid rheology and channel wall deformation, including obtaining new analytical expressions for the pressure drop reduction.

\begin{acknowledgments}
We thank E.\ Boyko for feedback on the manuscript and fruitful discussions on the use of the reciprocal theorem for weakly compliant channel flows.
\end{acknowledgments}

% we just have to keep switching references and references2 depending on who updates it last
\bibliography{references2,other_ref}

\end{document}